\documentclass[5p,final]{elsarticle}

\usepackage{lineno,hyperref}
\usepackage{natbib}
\usepackage{amsmath}

\usepackage{color}

\begin{document}

\begin{frontmatter}

\title{First principles phonon calculations in materials science}

\author[mymainaddress,mysecondaryaddress]{Atsushi Togo}
\ead{togo.atsushi@gmail.com}

\author[mymainaddress,mysecondaryaddress,mythirdaryaddress]{Isao Tanaka\corref{mycorrespondingauthor}}
\cortext[mycorrespondingauthor]{Corresponding author}
\ead{tanaka@cms.mtl.kyoto-u.ac.jp}

\address[mymainaddress]{Center for Elements Strategy Initiative for Structure Materials (ESISM), Kyoto University, Sakyo, Kyoto 606-8501, Japan}
\address[mysecondaryaddress]{Department of Materials Science and Engineering, Kyoto University, Sakyo, Kyoto 606-8501, Japan}
\address[mythirdaryaddress]{Nanostructures Research Laboratory, Japan Fine Ceramics Center, Atsuta, Nagoya 456-8587, Japan}

\begin{abstract}
 Phonon plays essential roles in dynamical behaviors and thermal
 properties, which are central topics in fundamental issues of materials
 science. The importance of first principles phonon calculations cannot
 be overly emphasized. {\tt Phonopy} is an open source code for such
 calculations launched by the present authors, which has been
 world-widely used.  Here we demonstrate phonon properties with
 fundamental equations and show examples how the phonon calculations are
 applied in materials science.
\end{abstract}

% \begin{keyword}
% \texttt{elsarticle.cls}\sep \LaTeX\sep Elsevier \sep template
% \MSC[2010] 00-01\sep  99-00
% \end{keyword}

\end{frontmatter}

\section{Introduction}
Application of first principles calculations in condensed matter physics
and materials science has greatly expanded when phonon calculations
became routine in the last decade. Thanks to the progress of high
performance computers and development of accurate and efficient density
functional theory (DFT) codes, a large set of first principles
calculations are now practical with the accuracy comparable to
experiments using ordinary PC clusters. In addition to electronic
structure information, a DFT calculation for solids provides energy and
stress of the system as well as the force on each atom. Equilibrium
crystal structures can be obtained by minimizing residual forces and
optimizing stress tensors. When an atom in a crystal is displaced from
its equilibrium position, the forces on all atoms in the crystal
raise. Analysis of the forces associated with a systematic set of
displacements provides a series of phonon frequencies. First principles
phonon calculations with a finite displacement method can be made in
this way. The present authors have launched a robust and easy-to-use
open-source code for first principles phonon calculations, {\tt
phonopy}~\cite{phonopy-project, phonopy, phonopy-QHA, Laurent-phph-2011,
phono3py, phonopy-evolution, phonopy-MSD, phonopy-CO2, Akamatsu-phonon,
Skelton-phonon-2014, Matsumoto-phonon, Edalati-phonon, Tamada-phonon,
Volker-phonon, Ikeda-phonon, Ankita-phonon}. The number of users is
rapidly growing world-wide, since the information of phonon is very
useful for accounting variety of properties and behavior of crystalline
materials, such as thermal properties, mechanical properties, phase
transition, and superconductivity. In this article, we show examples of
applications of the first principles phonon calculations.

In Sections
\ref{sec:harmonic-approximation}-\ref{sec:quasi-harmonic-approximation},
we take FCC-Al as examples of applications of
first principles phonon calculations.
%
% In practice, phonon and electronic structure calculations are well
% separated and these calculation details are presented as follows.
%
For the electronic structure
calculations, we employed the plane-wave basis projector augmented wave
method~\cite{PAW-Blochl-1994} in the framework of DFT within the
generalized gradient approximation in the Perdew-Burke-Ernzerhof
form~\cite{Perdew-PBE-1996} as implemented in the VASP
code~\cite{VASP-Kresse-1995,VASP-Kresse-1996,VASP-Kresse-1999}. A
plane-wave energy cutoff of 300 eV and an energy convergence criteria of
$10^{-8}$ eV were used. A $30\times 30\times 30$ $k$-point sampling mesh
was used for the unit cell and the equivalent density mesh was used for
the supercells together with a 0.2 eV smearing width of the
Methfessel-Paxton scheme~\cite{Methfessel-1989}.
For the phonon calculations, supercell and
finite displacement approaches were used with $3\times 3\times 3$
supercell of the conventional unit cell (108 atoms) and the atomic
displacement distance of 0.01 \AA. 

\section{Harmonic approximation}
\label{sec:harmonic-approximation} In crystals, it is presumed that atoms
move around their equilibrium positions $\mathbf{r}(l\kappa)$ with
displacements $\mathbf{u}(l\kappa)$, where $l$ and $\kappa$ are the
labels of unit cells and atoms in each unit cell, respectively. Crystal
potential energy $\Phi$ is presumed to be an analytic function of the
displacements of the atoms, and $\Phi$ is expanded as
% \begin{equation}
%  \Phi\left[\mathbf{u}(l_1\kappa_1),\ldots,\mathbf{u}(l_N\kappa_{n_\mathrm{a}})\right],
% \end{equation}
% where $N$ and $n_\mathrm{a}$ are the number of unit cells and the number
% of atoms, respectively. 
\begin{align}
 \label{eq:potential-expansion}
 \Phi & = \Phi_0  + \sum_{l\kappa}\sum_{\alpha} \Phi_\alpha(l\kappa)
 u_\alpha(l\kappa) \nonumber \\
 & + \frac{1}{2}\sum_{ll'\kappa\kappa'}
 \sum_{\alpha\beta}\Phi_{\alpha\beta}(l\kappa,l'\kappa')
 u_\alpha(l\kappa)u_\beta(l'\kappa') \nonumber \\
 & + \frac{1}{3!}\sum_{ll'l''\kappa\kappa'\kappa''}
 \sum_{\alpha\beta\gamma} \nonumber \\
 & \;\;\;\;\;\;\Phi_{\alpha\beta\gamma}(l\kappa,l'\kappa',l''\kappa'')
 u_\alpha(l\kappa) u_\beta(l'\kappa') u_\gamma(l''\kappa'') + \cdots
\end{align}
where $\alpha$, $\beta$, $\cdots$ are the Cartesian indices. The
coefficients of the series expansion, $\Phi_0$, $\Phi_\alpha(l\kappa)$,
$\Phi_{\alpha\beta}(l\kappa,l'\kappa')$, and,
$\Phi_{\alpha\beta\gamma}(l\kappa,l'\kappa',l''\kappa'')$, are the
zero-th, first, second, and third order force constants,
respectively. With small displacements at constant volume, the problem
of atomic vibrations is solved with the second-order terms as the
harmonic approximation, and the higher order terms are treated by
the perturbation theory.

With a force $F_\alpha(l\kappa) = -\frac{\partial \Phi}{\partial
u_\alpha(l\kappa)}$, an element of second-order force constants
$\Phi_{\alpha\beta}(l\kappa,l'\kappa')$ is obtained by
\begin{equation}
 \Phi_{\alpha\beta}(l\kappa,l'\kappa') = \frac{\partial^2\Phi}
  {\partial u_\alpha(l\kappa)
  \partial u_\beta(l'\kappa')} = -\frac{\partial
  F_\beta(l'\kappa')}{\partial u_\alpha(l\kappa)}.
\end{equation}
%
% In the finite displacement approach, the equation for the force
% constants is approximated as
% \begin{equation}
%  \Phi_{\alpha\beta}(l\kappa, l'\kappa') \simeq -\frac{
%   F_\beta\left[l'\kappa';\Delta u_\alpha{(l\kappa)}\right]} {\Delta
%   u_\alpha(l\kappa)},
% \end{equation}
% where $F_\beta\left[l'\kappa';\Delta r_\alpha{(l\kappa)}\right]$ is the
% force on the atom $l'\kappa'$ with a small finite displacement $\Delta
% u_\alpha(l\kappa)$.
%
Crystal symmetry is utilized to improve the numerial
accuracy of the force constants and to reduce the computational
cost. The more details on the calculation of force constants are found
in Ref.~\cite{Laurent-phph-2011,phono3py}.

As found in text books\cite{Electrons-and-phonons,
Physics-of-phonons, Thermodynamics-of-crystals,
IntroductionToLatticeDynamics}, dynamical property of atoms in the
harmonic approximation is obtained by solving eigenvalue problem of
dynamical matrix $\mathrm{D}(\mathbf{q})$,
\begin{align}
 \label{eq:eigenvalue-problem}
 \mathrm{D}(\mathbf{q}) \mathbf{e}_{\mathbf{q}j} =
 \omega_{\mathbf{q}j}^2 \mathbf{e}_{\mathbf{q}j},\;\; \text{or}\;\;
 \sum_{\beta\kappa'} D^{\alpha\beta}_{\kappa\kappa'}(\mathbf{q})
  e^{\beta\kappa'}_{\mathbf{q}j} =
 \omega_{\mathbf{q}j}^2 e^{\alpha\kappa}_{\mathbf{q}j}
\end{align}
with
\begin{align}
 \label{eq:dynamical-matrix}
 D^{\alpha\beta}_{\kappa\kappa'}(\mathbf{q}) = 
 \sum_{l'} \frac{\Phi_{\alpha\beta}(0\kappa, l'\kappa')}{\sqrt{m_\kappa m_{\kappa'}}}
 e^{i\mathbf{q}\cdot[\mathbf{r}(l'\kappa')-\mathbf{r}(0\kappa)]},
\end{align}
where $m_\kappa$ is the mass of the atom $\kappa$, $\mathbf{q}$ is the
wave vector, and $j$ is the band index. $\omega_{\mathbf{q}j}$ and
$\mathbf{e}_{\mathbf{q}j}$ give the phonon frequency and polarization
vector of the phonon mode labeled by a set $\{\mathbf{q}, j\}$,
respectively. Since $\mathrm{D}(\mathbf{q})$ is an Hermitian matrix, its
eigenvalues, $\omega_{\mathbf{q}j}^2$, are real.
%
% Each eigensolution corresponds to a
% phonon mode that is labeled by a set $\{\mathbf{q}, j\}$.
%
Usually
$\mathrm{D}(\mathbf{q})$ is arranged to be a $3n_\mathrm{a} \times
3n_\mathrm{a}$ matrix~\cite{IntroductionToLatticeDynamics}, where 3
comes from the freedom of the Cartesian indices for crystal and
$n_\mathrm{a}$ is the nubmer of atoms in a unit cell. Then
$\mathbf{e}_{\mathbf{q}j}$ becomes a complex column vector with
$3n_\mathrm{a}$ elements, and usually $\mathbf{e}_{\mathbf{q}j}$ is
normalized to be 1, i.e.,
$\sum_{\alpha\kappa}|e^{\alpha\kappa}_{\mathbf{q}j}|^2=1$.
$\mathbf{e}_{\mathbf{q}j}$ contains information of collective motion of
atoms. This may be understood as a set of atomic displacement vectors,
\begin{align}
\label{eq:corrective-displacements}
\left\{ \right.
& \mathbf{u}(l1), \ldots, \mathbf{u}(l\kappa)
\left. \right\} =\nonumber \\
& \left\{
\frac{A}{\sqrt{m_1}}
\mathbf{e}^1_{\mathbf{q}j}
e^{i\mathbf{q}\cdot\mathbf{r}(l1)}, 
\ldots,
\frac{A}{\sqrt{m_{n_\mathrm{a}}}}
\mathbf{e}^{n_\mathrm{a}}_{\mathbf{q}j} e^{i\mathbf{q}\cdot\mathbf{r}(l\kappa)}
\right\},
\end{align}
where $A$ is the complex constant undetermined by
Eq.~(\ref{eq:eigenvalue-problem}), and
${\mathbf{e}^{\kappa}_{\mathbf{q}j}}^\mathrm{T}=(e^{x\kappa}_{\mathbf{q}j},
e^{y\kappa}_{\mathbf{q}j}, e^{z\kappa}_{\mathbf{q}j})$.

\begin{figure}[ht]
 \begin{center}
  \includegraphics[width=0.80\linewidth]{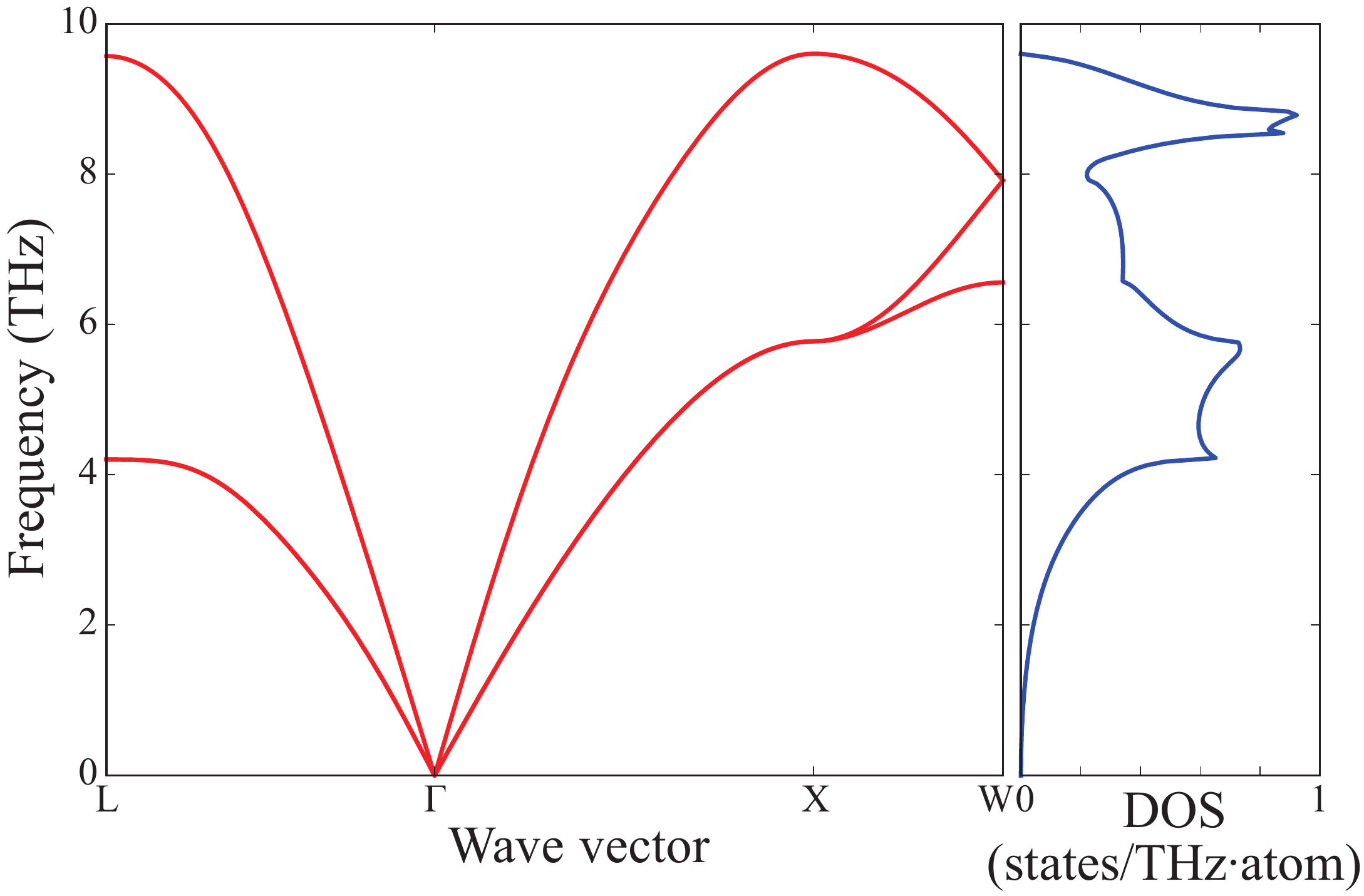}
  \caption{(color online) \label{fig:Al-phonon}
  Phonon band structure and DOS of Al.}
 \end{center}
\end{figure}

As a typical example, the phonon band structure and phonon density of
states (DOS) of Al are shown in
Fig.~\ref{fig:Al-phonon}. The phonon DOS is defined as
\begin{align}
g(\omega) = \frac{1}{N}\sum_{\mathbf{q}j} \delta(\omega - \omega_{\mathbf{q}j}),
\end{align} 
where $N$ is the number of unit cells in crystal. Divided by $N$,
$g(\omega)$ is normalized so that the integral over frequency becomes
$3n_\mathrm{a}$.
% This integration is achieved usually by smearing method
% or tetrahedron
% method\cite{MacDonald-tetrahedron-1979,Blochl-tetrahedron-1994}.
The phonon band structure can be directly comparable with experimental
data by neutron or X-ray inelastic scattering. They often show
reasonable agreements~\cite{Ankita-phonon, Koermann-phonon,
Koermann-phonon2}. Frequency data by Raman and infrared (IR)
spectroscopy can also be well reproduced~\cite{phonopy-CO2,
kuwabara-phonon}.  Irreducible representations of phonon modes, which
can be used to assign Raman or IR active modes, are calculated from
polarization vectors~\cite{phonopy-CO2, Dynamics-of-perfect-crystals}.
Atom specific phonon DOS projected along a unit direction vector
$\hat{\mathbf{n}}$ is defined as
\begin{align}
\label{eq:partial-DOS}
 g_\kappa(\omega,\hat{\mathbf{n}}) = \frac{1}{N}\sum_{\mathbf{q}j} \delta(\omega -
  \omega_{\mathbf{q}j}) |\hat{\mathbf{n}} \cdot
  \mathbf{e}^\kappa_{\mathbf{q}j}|^2.
\end{align}
This $g_\kappa(\omega,\hat{\mathbf{n}})$ can be directly compared with that
measured by means of nuclear-resonant inelastic scattering using
synchrotron radiation. In Ref.~\cite{Tamada-phonon}, phonon calculations
of L1$_0$-type FePt projected along the $c$-axis and basal plane are well
comparable to experimental $^{57}$Fe
nuclear-resonant inelastic scattering spectra measured at 10 K in the
parallel and perpendicular geometries, respectively.

\begin{figure}[ht]
 \begin{center}
  \includegraphics[width=0.80\linewidth]{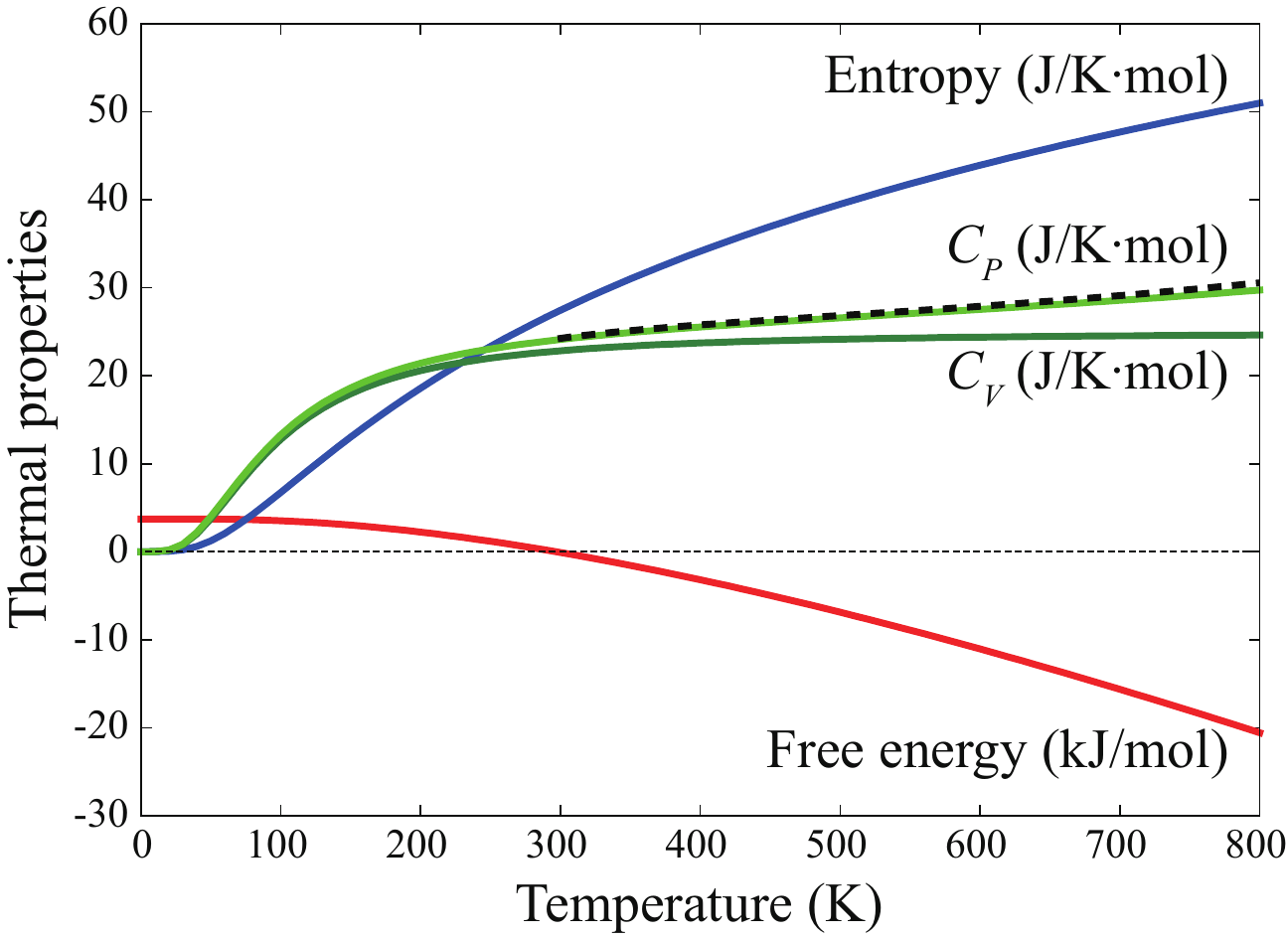}
  \caption{(color online) \label{fig:Al-thermal-properties} Thermal
  properties of Al. Entropy, $C_V$, and Helmholtz free energy were
  calculated with harmonic approximation
  (Sec.~\ref{sec:harmonic-approximation}). QHA was employed to obtain
  $C_P$ (Sec.~\ref{sec:quasi-harmonic-approximation}). Physical units
  are shown with labels of the physical properties, and the value of the
  vertical axis is shared by them. Dotted curve depicts the experiment
  of $C_P$~\cite{chase1998nist}.}
 \end{center}
\end{figure}

Once phonon frequencies over Brillouin zone are known, from the
canonical distribution in statistical mechanics for phonons under the
harmonic approximation, the energy $E$ of phonon system is given as
\begin{align}
\label{eq:internal-energy}
 E = \sum_{\mathbf{q}j}\hbar\omega_{\mathbf{q}j}\left[\frac{1}{2} +
 \frac{1}{\exp({\hbar\omega_{\mathbf{q}j}/\mathrm{k_B}T})-1}\right],
\end{align}
% \begin{align}
% \label{eq:phonon-occupation-number}
%  n_{\mathbf{q}j} =
%  \left[\exp({\hbar\omega_{\mathbf{q}j}/\mathrm{k_B}T})-1\right]^{-1},
% \end{align}
where $T$, $k_\mathrm{B}$, and $\hbar$ are the temperature, the
Boltzmann constant, and the reduced Planck constant, respectively. Using
the thermodynamic relations, a number of thermal properties, such as
constant volume heat capacity $C_V$, Helmholtz free energy $F$, and
entropy $S$, can be computed as functions of
temperature~\cite{IntroductionToLatticeDynamics}:
\begin{align}
\label{eq:heat-capacity}
 C_V = \sum_{\mathbf{q}j} C_{\mathbf{q}j} = \sum_{\mathbf{q}j} k_\mathrm{B}
 \left(\frac{\hbar\omega_{\mathbf{q}j}}{k_\mathrm{B} T} \right)^2
 \frac{\exp(\hbar\omega_{\mathbf{q}j}/k_\mathrm{B}
 T)}{[\exp(\hbar\omega_{\mathbf{q}j}/k_\mathrm{B} T)-1]^2},
\end{align}
\begin{align}
\label{eq:helmholtz-free-energy}
 F = \frac{1}{2} \sum_{\mathbf{q}j} \hbar\omega_{\mathbf{q}j}
 + k_\mathrm{B} T \sum_{\mathbf{q}j} \ln \bigl[1
 -\exp(-\hbar\omega_{\mathbf{q}j}/k_\mathrm{B} T) \bigr],
\end{align}
and
\begin{align}
\label{eq:entropy}
 S = & \frac{1}{2T} \sum_{\mathbf{q}j} \hbar\omega_{\mathbf{q}j}
 \coth\left[\hbar\omega_{\mathbf{q}j}/2k_\mathrm{B}T\right] \nonumber \\
 & - k_\mathrm{B}
 \sum_{\mathbf{q}j}
 \ln\left[2\sinh(\hbar\omega_{\mathbf{q}j}/2k_\mathrm{B}T)\right].
\end{align}
The calculated $F$, $C_V$, and $S$ for Al are shown in
Fig.~\ref{fig:Al-thermal-properties}.

\section{Mean square atomic displacements}
\label{sec:mean-square-displacement}
With the phase factor convention of the dynamical matrix used in
Eq.~(\ref{eq:dynamical-matrix}), an atomic displacement operator
is written as,
\begin{align}
 \label{eq:displacement-operator}
 \hat{u}_\alpha(l\kappa)= 
 \sqrt{ \frac{\hbar}{2Nm_\kappa} }
 \sum_{\mathbf{q}j}
 \frac{\hat{a}_{\mathbf{q}j}+\hat{a}^\dagger_{-\mathbf{q}j}}
 {\sqrt{\omega_{\mathbf{q}j}}}
 e^{\alpha\kappa}_{\mathbf{q}j}
 e^{i\mathbf{q}\cdot\mathbf{r}(l\kappa)},
\end{align}
where $\hat{a}^\dagger$ and $\hat{a}$ are the creation and annihilation
operators, respectively.  Distributions of atoms around their
equilibrium positions are then obtained as the expectation values of
Eq.~(\ref{eq:displacement-operator}). The mean square atomic
displacement projected along $\hat{\mathbf{n}}$ is obtained as
% \begin{align}
%  \left\langle |\hat{u}_\alpha(\kappa)|^2 \right\rangle =
%  \frac{\hbar}{2Nm_\kappa} \sum_{\mathbf{q}j}
%  \frac{1+2n_{\mathbf{q}j}}{\omega_{\mathbf{q}j}}
%  |e^{\alpha\kappa}_{\mathbf{q}j}|^2.
% \end{align}
\begin{align}
 \label{eq:mean-square-displacement}
 \left\langle |\hat{u}_{\hat{\mathbf{n}}}(\kappa)|^2 \right\rangle =
 \frac{\hbar}{2Nm_\kappa} \sum_{\mathbf{q}j}
 \frac{1+2n_{\mathbf{q}j}}{\omega_{\mathbf{q}j}}
 |\hat{\mathbf{n}}\cdot\mathbf{e}^{\kappa}_{\mathbf{q}j}|^2.
\end{align}
Eq.~(\ref{eq:mean-square-displacement}) is lattice-point ($l$)
independent since the phase factor disappears.
Anisotropic atomic displacement parameters (ADPs) to estimate the atom
positions during thermal motion can also be computed and compared with
experimental neutron diffraction data.  Thermal ellipsoids may be
discussed using mean square displacement matrix $\mathrm{B}(\kappa)$
defined by
\begin{align}
 \label{eq:mean-square-displacement-matrix}
 \mathrm{B}(\kappa) = \frac{\hbar}{2Nm_\kappa}
 \sum_{\mathbf{q}j}
 \frac{1+2n_{\mathbf{q}j}}{\omega_{\mathbf{q}j}}
 \mathbf{e}^{\kappa}_{\mathbf{q}j} \otimes \mathbf{e}^{\kappa*}_{\mathbf{q}j}.
\end{align}
The shape and orientation of an ellipsoid is obtained solving eigenvalue
problem of this matrix. The method has been applied to show the ORTEP
(Oak Ridge Thermal Ellipsoid Plot)-style drawing of
ADPs~\cite{Volker-phonon}. Ref.~\cite{phonopy-MSD} shows an example for
a ternary carbide Ti$_3$SiC$_2$ having a layered structure known as MAX
phases, in which we can see good agreement between 
calculated and experimental ADPs.
%  that the calculation reproduces well the
% experimentally obtained ADPs.

\section{Quasi-harmonic approximation}
\label{sec:quasi-harmonic-approximation}

% Phonon frequency is necessary to be modfied when anharmonic terms are
% not negligible compared to the harmonic terms in the Taylor series
% expansion of potential energy.  Phonon renomalization theory and
% self-consistent phonon theory were
% developed\cite{Thermodynamics-of-crystals,
% IntroductionToLatticeDynamics} for this purpose. However they are not
% often used since they are computationally demanding to achieve.

\begin{figure}[ht]
 \begin{center}
  \includegraphics[width=1.00\linewidth]{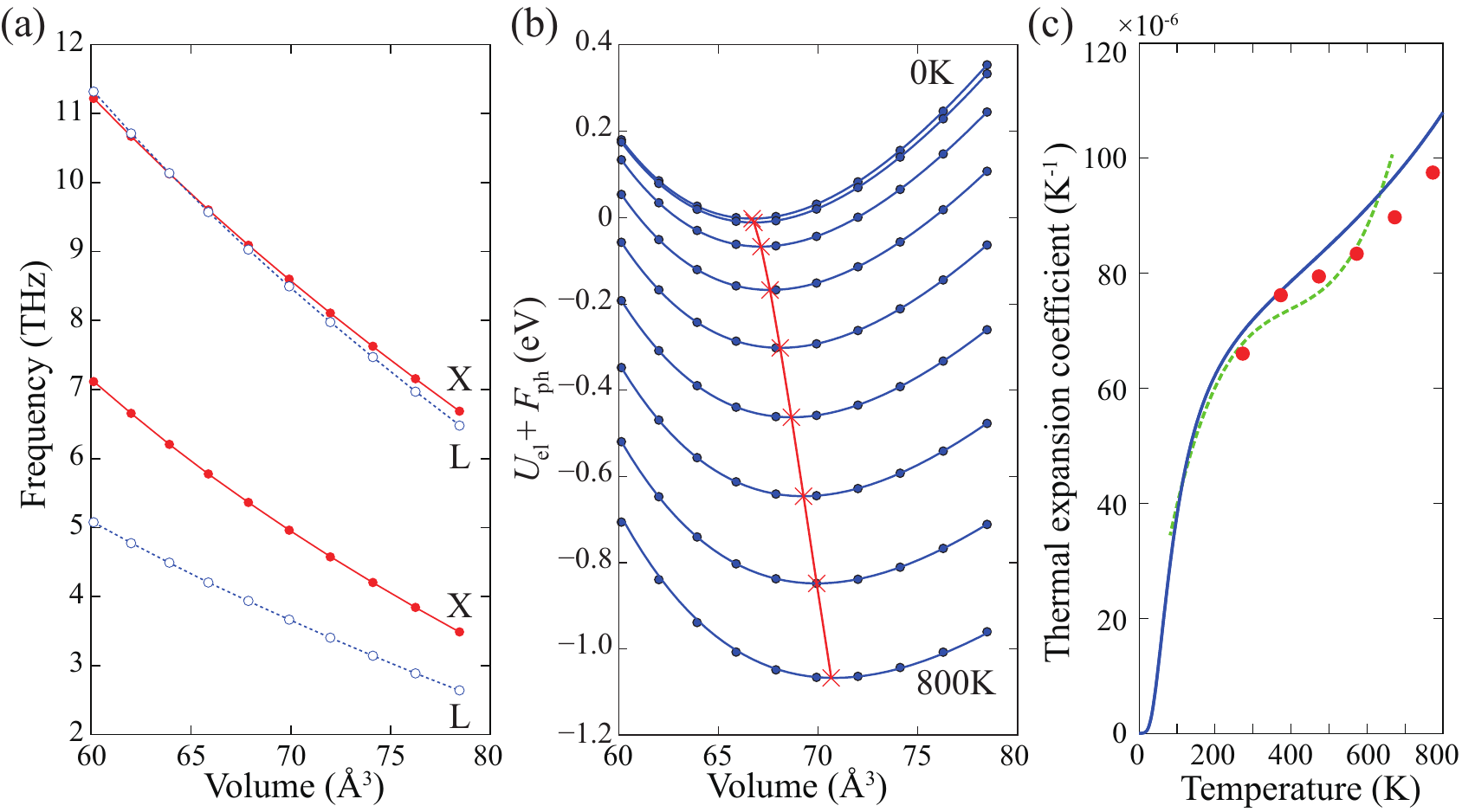} \caption{(color
  online) \label{fig:Al-QHA} (a) Phonon frequencies of Al at $X$ and $L$
  points with respect to unit cell volume are shown by filled and open
  circles, respectively. The solid and dotted lines are guides to the
  eye. (b) $U_\mathrm{el} + F_\mathrm{ph}$ of Al with respect to volume
  at temperatures from 0 to 800 K with 100 K step are depicted by filled
  circles and the values are fit by the solid curves. Cross symbols show
  the energy bottoms of the respective curves and simultaneously the
  equilibrium volumes. Lines connecting the cross symbols are guides to
  the eye. (c) Volumetric
  thermal expansion coefficient of Al. Calculation is shown with solid
  curve and experiments are depicted by filled circle
  symbols~\cite{Al-Wilson-1941} and dotted curve~\cite{Nix-Al-1941}.}
 \end{center}
\end{figure}

By changing volume, phonon properties vary since the crystal
potential is an anharmonic function of volume.
%
% This is also an
% anharmonic effect however anharmonicity is treated implicitly.
%
In this article, the term ``quasi-harmonic approximation (QHA)`` means
this volume dependence of phonon properties, but the harmonic
approximation is simply applied at each volume.
Figure \ref{fig:Al-QHA}a shows calculated phonon frequencies of Al at
$X$ and $L$ points with respect to ten different unit-cell
volumes. Typically phonon frequency decreases by increasing volume, and
the slope of each phonon mode is nearly constant in the wide volume
range.
The normalized slope is called mode-Gr\"{u}neisen parameter
that is defined as
\begin{align}
 \gamma_{\mathbf{q}j}(V) = -\frac{V}{\omega_{\mathbf{q}j}(V)}
 \frac{\partial \omega_{\mathbf{q}j}(V)}{\partial V}.
\end{align}
Once dynamical matrix is known, $\gamma_{\mathbf{q}j}$ is easily
calculated from the volume derivative of the dynamical
matrix\cite{Thermodynamics-of-crystals},
\begin{align}
\label{eq:Gruneisen-perturbation}
 \gamma_{\mathbf{q}j}(V)
 =-\frac{V}{2(\omega_{\mathbf{q}j})^2}
 \sum_{\alpha\beta\kappa\kappa'}e^{\alpha\kappa*}_{\mathbf{q}j}
 \frac{\partial D^{\alpha\beta}_{\kappa\kappa'}(V,\mathbf{q})}
 {\partial V}e^{\beta\kappa'}_{\mathbf{q}j}.
\end{align}
The quantity can be related to macroscopic Gr\"{u}neisen parameter
$\gamma$ using mode contributions to the heat capacity $C_{\mathbf{q}j}$
found in Eq.~(\ref{eq:heat-capacity}), $\gamma=\sum_{\mathbf{q}j}
\gamma_{\mathbf{q}j}C_{\mathbf{q}j}/C_V$ \cite{Physics-of-phonons,
IntroductionToLatticeDynamics}.

Silicon is known as a famous exception to have large negative
mode-Gr\"{u}neisen parameters. Mode-Gr\"{u}neisen parameter is a measure
of anharmonicity of phonon modes and is related to third-order force
constants directly~\cite{Thermodynamics-of-crystals}. Therefore crystals
that possess large anharmonic terms beyond third-order terms in
Eq.~(\ref{eq:potential-expansion}) can show non-linear change of phonon
frequency with respect to volume. This is often observed in crystals
that exhibit second- or higher-order structural phase
transitions~\cite{phonopy}.

% The mode-Gr\"{u}neisen parameter may be considered as the coefficient of the
% first-order term in the Taylor series of $\omega_{\mathbf{q}j}(V)$,
% \begin{align}
% \frac{\omega_{\mathbf{q}j}(V)}{\omega_{\mathbf{q}j}(V_0) } = 
% 1 -
%  \gamma_{\mathbf{q}j}(V_0)\epsilon
%  - \gamma'_{\mathbf{q}j}(V_0)\epsilon^2 
%  + \cdots,
% \end{align}
% where $\epsilon = \frac{V - V_0}{V_0}$.

% Figure 6 shows calculated mode Gr\"{u}neisen parameter for Si with diamond
% structure. In Fig. 6(a), neighboring q-points in each band segment are
% connected considering their phonon symmetry to treat band crossing
% correctly. In the plot, the colors of phonon bands correspond to those
% of mode-Gr\"{u}neisen parameters shown in Fig. 6(b). Si with diamond
% structure is a famous example showing negative mode-Gr\"{u}neisen
% parameters.

The phonon frequency influences the phonon energy
through Eq.~(\ref{eq:internal-energy}). The thermal properties
are thereby affected.
Using thermodynamics definition, thermodynamic variables at constant
volume is transformed to those at constant pressure that is often more
easily measurable in experiments. Gibbs free energy $G(T, p)$ at given
temperature $T$ and pressure $p$ is obtained from Helmholtz free energy
$F(T;\,V)$ through the transformation,
\begin{equation}
\label{eq:gibbs-free-energy}
 G(T, p) = \min_V \left[F(T;\,V) + pV \right],
\end{equation}
where the right hand side of this equation means finding a minimum value
in the square bracket by changing volume $V$. We may approximate
$F(T;\,V)$ by the sum of electronic internal energy $U_\mathrm{el}(V)$
and phonon Helmholtz free energy $F_\mathrm{ph}(T;\,V)$, i.e., $F(T;\,V)
\simeq U_\mathrm{el}(V) + F_\mathrm{ph}(T;\,V)$. $U_\mathrm{el}(V)$ is
obtained as total energy of electronic structure from the first
principles calculation, and the first principles phonon calculation at
$T$ and $V$ gives $F_\mathrm{ph}(T;\,V)$. The calculated
$U_\mathrm{el}(V) + F_\mathrm{ph}(T;\,V)$ are depicted by the filled
circle symbols in Fig.~\ref{fig:Al-QHA}b, where the ten volume points
chosen are the same as those in Fig.~\ref{fig:Al-QHA}a. The nine curves
are the fits to equation of states (EOS) at temperatures from 0 to 800 K
with 100 K step. Here the Vinet EOS~\cite{Vinet-1989} was used to fit
the points to the curves though any other functions can be used for the
fitting.
%  if minimum values in Eq.~(\ref{eq:gibbs-free-energy}) are found
% well.
 The minimum values at the temperatures are depicted by the cross
symbols in Fig.~\ref{fig:Al-QHA}b and are the Gibbs free energies at the
temperatures and the respective equilibrium volumes are simultaneously
given. Volumetric thermal expansion coefficient,
$\beta(T)=\frac{1}{V(T)}\frac{\partial V(T)}{\partial T}$, is obtained
from the calculated equilibrium volumes $V(T)$ at dense temperature
points. $\beta(T)$ for Al is shown in Fig.~\ref{fig:Al-QHA}c, where we
can see that the calculation shows reasonable agreements with the
experiments. In thermodynamics, heat capacity at constant pressure $C_P$
is given by
\begin{align}
 \label{eq:Cp}
  C_P(T, p) & = -T\frac{\partial^2 G(T, p)}{\partial T^2} \nonumber \\
  = C_V(T, &V(T, p)) + T\frac{\partial V(T, p)}{\partial T} \frac{\partial S(T; V)}{\partial
  V} \biggr|_{V=V(T, p)}.
\end{align}
In Eq.~(\ref{eq:Cp}), the second term of the second equation is
understood as the contribution to heat capacity from thermal
expansion. $C_P$ for Al is shown in
Fig.~\ref{fig:Al-thermal-properties}. The agreement of the calculation
with the experiment is excellent. At high temperatures, the difference
between $C_P$ and $C_V$ is not negligible in Al. Therefore it is
essential to consider thermal expansion for heat capacity.

% \begin{figure}[ht]
%  \begin{center}
%   \includegraphics[width=0.70\linewidth]{Al-thermal_expansion.pdf}
%   \caption{(color online) \label{fig:Al-thermal-expansion} Volumetric
%   thermal expansion coefficient of Al. Calculation is shown with solid
%   curve and experiments are depicted by filled circle
%   symbols~\cite{Al-Wilson-1941} and dotted curve~\cite{Nix-Al-1941}.}
%  \end{center}
% \end{figure}

QHA is known as a reasonable approximation in a wide temperature range below
melting point except for temperatures very close to melting point where
higher-order terms in Eq.~(\ref{eq:potential-expansion}) become
evident~\cite{Grabowski-2009}.

\section{Stability condition and imaginary mode}
At equilibrium, $\frac{\partial \Phi}{\partial r_\alpha(l\kappa)}=0$, a
crystal is dynamically (mechanically) stable if its potential energy
always increases against any combinations of atomic displacements. In
the harmonic approximation, this is equivalent to the condition that all
phonons have real and positive
frequencies\cite{Thermodynamics-of-crystals}. However under virtual
thermodynamic conditions, imaginary frequency or negative eigenvalue can
appear in the solution of Eq.~(\ref{eq:eigenvalue-problem}). This
indicates dynamical instability of the system, which means that the
corrective atomic displacements of
Eq.~(\ref{eq:corrective-displacements}) reduce the potential energy in
the vicinity of the equilibrium atomic positions.

\begin{figure}[ht]
 \begin{center}
  \includegraphics[width=0.9\linewidth]{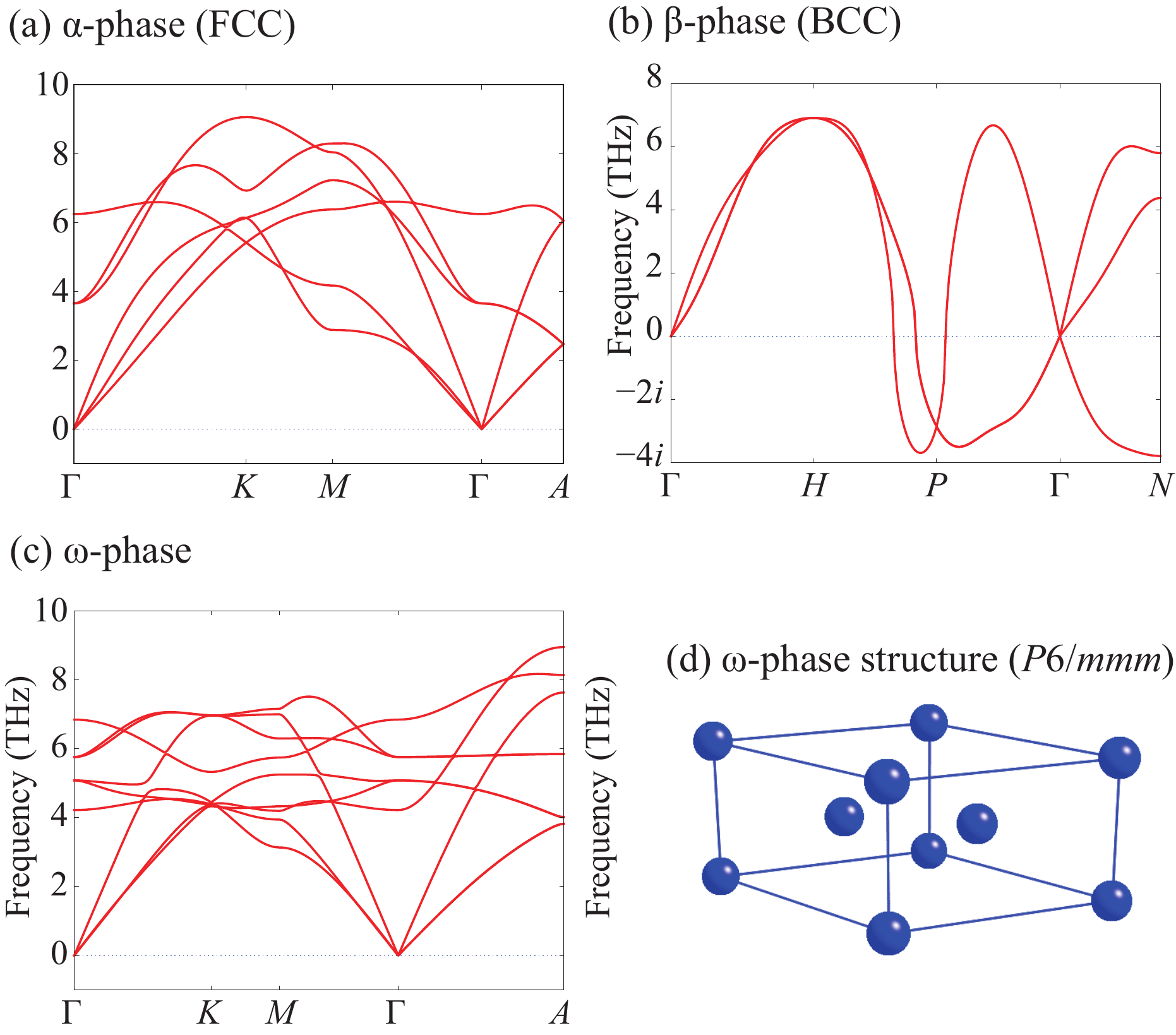}
  \caption{(color online) \label{fig:Ti-phonon-bands} Phonon band
  structures of (a) $\alpha$-Ti (HCP), (b) $\beta$-Ti (BCC), and (c)
  $\omega$-Ti. 
  The figure (d) shows the hexagonal crystal structure of
  $\omega$-Ti.
}
 \end{center}
\end{figure}

Imaginary mode provides useful information to study displacive phase
transition. A typical example is shown in
Figs.~\ref{fig:Ti-phonon-bands}a to
\ref{fig:Ti-phonon-bands}c~\cite{Edalati-phonon}.
%
% Structures of all three phases, i.e., $\omega$,
% $\alpha$, and $\beta$, are optimized prior to the phonon
% calculations.
%
Imaginary modes can be found only for $\beta$-Ti, that has BCC
structure, at both $P$ and $N$ points. This indicates that $\beta$-Ti is
unstable at low temperature. Such imaginary modes cannot be seen for
either $\omega$-Ti whose crystal structure is shown in
Fig.~\ref{fig:Ti-phonon-bands}d or $\alpha$-Ti (HCP).  Experimentally
$\beta$-Ti is known to occur above 1155K. At such high temperatures,
large atomic displacements can stabilize the BCC structure. In such a
case, the perturbation approach is invalid. Phonons with large atomic
displacements may be treated by self-consistent phonon
method\cite{Thermodynamics-of-crystals,
Errea-self-consistent-phonon-2014} or by a combination of molecular
dynamics and lattice dynamics calculation\cite{Wang-MD-1990,
Lee-MD-1993, Sun-MD-2014}, which is not discussed in this article.

\begin{figure}[ht]
 \begin{center}
  \includegraphics[width=0.9\linewidth]{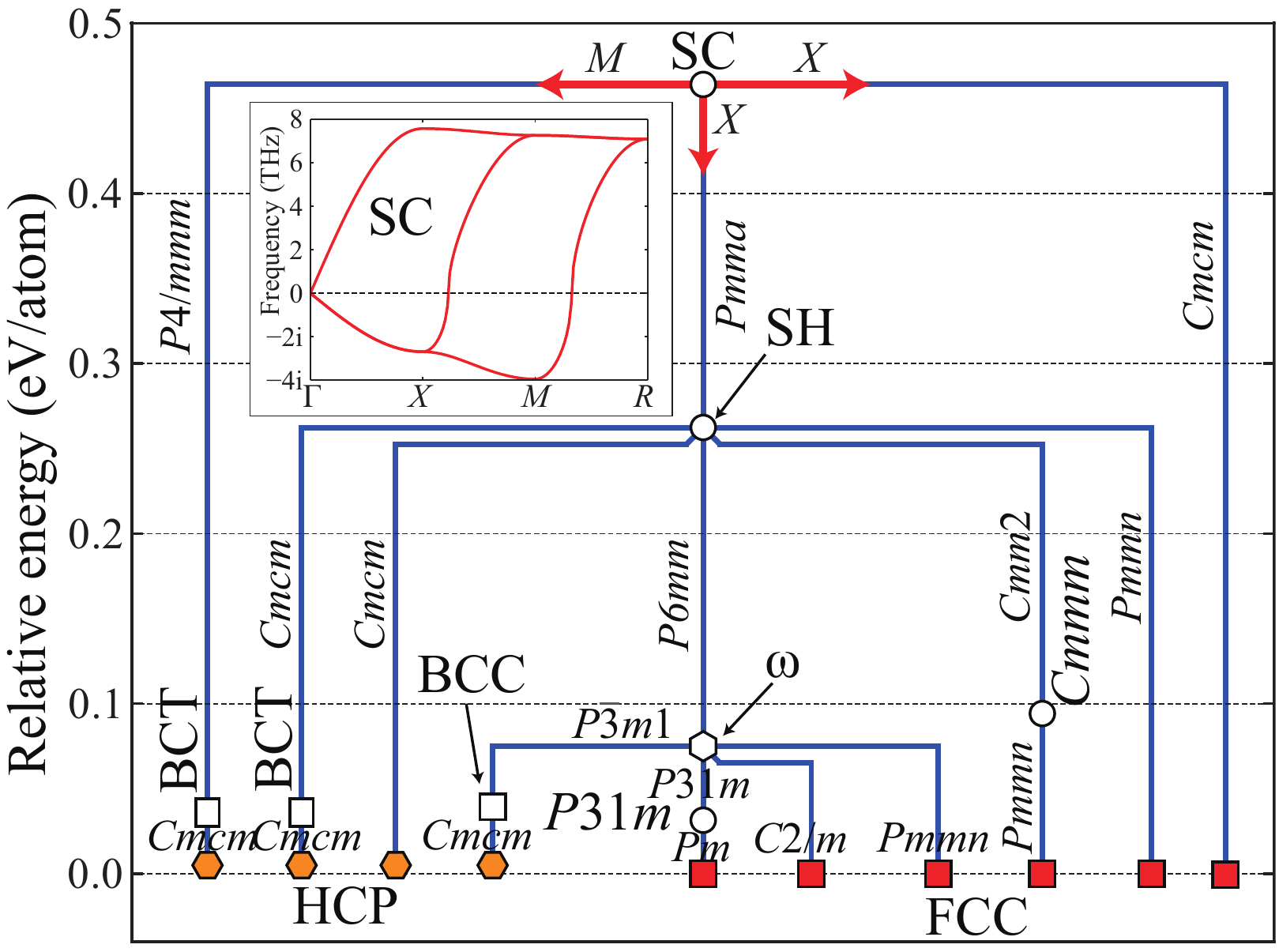} \caption{(color
  online) \label{fig:Cu-tree} Line diagram of structural transition
  pathways in Cu. The inset shows phonon band structure of simple cubic
  (SC) Cu. Open and filled symbols represent dynamically unstable and
  stable crystal structures, respectively. Lines connecting these
  symbols are the phase transition pathways for which space-group types
  are shown near the lines. }
 \end{center}
\end{figure}

A given structure having imaginary phonon modes can be led to
alternative structures through continuous atomic displacements and
lattice deformations. The present authors systematically investigated
the evolution of crystal structures from the simple cubic (SC)
structure~\cite{phonopy-evolution}. The inset of Fig.~\ref{fig:Cu-tree}
shows the phonon band structure of SC-Cu ($Pm\bar{3}m$). Imaginary modes
can be found at $M(1/2, 1/2, 0)$ and $X(1/2, 0, 0)$ points. Then the SC
structure is deformed along these directions. In order to accommodate
the deformation in the calculation model with the periodic boundary
condition, the unit cells are expanded by $2\times 2\times 1$ for the
$M$ point and $2\times 1\times 1$ for the $X$ point.  The $M$ point
deformation breaks the crystal symmetry of SC ($Pm\bar{3}m$) to
$P4/nmm$. The doubly degenerated instability at the $X$ point leads to
$Pmma$ and $Cmcm$ as highest possible symmetries. The deformed crystal
structures are relaxed by first principles calculations imposing the
corresponding space-group operations. After these procedures,
body-centered tetragonal (BCT), simple hexagonal (SH), and FCC are
respectively formed. The whole procedure finishes when all crystal
structures at the end-points are found to be dynamically stable. Finally
a treelike structure of crystal structure relationships was drawn as
shown in Fig.~\ref{fig:Cu-tree}, where thick lines indicate phase
transition pathways (PTPs). The space-group type written near a line is
a common subgroup of initial and final structures. The presence of the
line indicates that the energy decreases monotonically with the phase
transition. In other words, the transition can take place spontaneously
without any energy barrier. The line is terminated when the final
structure is dynamically stable.

In the line diagram, $\omega$ is located at the junction of two
pathways, i.e., $\omega\rightarrow$ BCC $\rightarrow$ HCP and $\omega
\rightarrow$ FCC. The instability of $\omega$ at the $\Gamma$ point
leads to BCC, which is still dynamically unstable and eventually leads
to HCP. Another instability at the $M$ point leads to FCC. The other
instability at the $K$ point, which is doubly degenerate, leads to
FCC. On the path from $\omega$ to BCC, the crystal symmetry of $\omega$
having the space-group type of $P6/mmm$ is once lowered to $P\bar{3}m1$
and then becomes $Im\bar{3}m$ (BCC) after the geometry
optimization. Both $\omega$- and BCC-Cu are dynamically unstable, which
can be formed only under crystal symmetry constraints. FCC-Cu is, of
course, dynamically stable. Several PTPs between BCC and FCC have been
proposed in literature. However, they are mostly based only upon
investigation of continuous lattice deformation. For example in the
classical Bain path, formation of BCT in between BCC and FCC
is considered. Formation of SC is taken into account in the ``trigonal
Bain path.'' Normal modes of phonon, which should be most adequate to
describe the collective atomic displacements, have not been
considered. The presence of $\omega$ as the lowest energy barrier in the
BCC-FCC pathway had not been reported before
Ref.~\cite{phonopy-evolution}. The situation is the same for the BCC-HCP
transition, known as the Burgers path. The Burgers path was thought to
be quite complicated from the viewpoint of the lattice continuity. On
the basis of the present study, it can be easily pointed out that the
BCC-HCP transition pathway is along the space-group type of $Cmcm$.

% The evolution diagram for Ti is shown in Fig. 9(b). Compared with the
% diagram of Cu in Fig. 9(a), the path connecting SH and $\omega$
% disappears in Ti. There is an energy barrier between SH and $\omega$ in
% Ti. A separate calculation finds that $\omega$-Ti is dynamically stable
% with the energy lower than HCP. $\omega$-Ti is experimentally reported
% as a stable phase under external pressure of 2 GPa at 293 K. The
% critical pressure decreases with decreasing temperature, suggesting that
% $\omega$ is more stable than HCP at temperatures below 90 K. However,
% the HCP $\rightarrow \omega$ phase transition under the ordinary
% pressure has not yet been reported experimentally. The transition under
% pressure showed a large hysteresis. These experimental results imply the
% presence of a large energy barrier for the transition. The present
% result is consistent to such experimental results.

Evolution diagram was constructed in the same way for Na$R$TiO$_4$ ($R$:
rare-earth metal) with Ruddlesden-Popper type structure
\cite{Akamatsu-phonon}. Inversion symmetry breaking by oxygen octahedral
rotations was unambiguously demonstrated. The mechanism is expected to
lead to many more families of acentric oxides.

\section{Interactions among phonons and lattice thermal conductivity} 

Using the harmonic phonon coordinates, anharmonic terms in
Eq.~(\ref{eq:potential-expansion}) are transformed to a picture of
phonon-phonon interactions~\cite{Laurent-phph-2011,
Physics-of-phonons}. Lattice thermal conductivity can be accurately
calculated by solving linearized Boltzmann transport equation with the
phonon-phonon interaction strength obtained using first principles
calculation~\cite{phono3py, Laurent-LBTE-2013,
seko-informatics}. Although the computational cost for such calculations
is many orders of magnitudes higher than the ordinary DFT calculations
of primitive cells, such calculations have already been applied for many
simple compounds and computed lattice thermal conductivities show good
agreements with experimental data~\cite{phono3py,
seko-informatics}. Calculations with special focus on searching
thermoelectric materials have also been made~\cite{Skelton-phonon-2014,
Ankita-phonon, seko-informatics}.
% \cite{Jonathan-Raman-2015}

\section*{Acknowledgements}
The research leading to these results was supported by Grant-in-Aid for
Scientific Research on Innovative Areas ``Nano Informatics'' (Grant
No. 25106005) and for Young Scientists (B) (Grant No. 26820284) both
from JSPS.  Support from MEXT through ESISM is also acknowledged.

\bibliography{phonon_viewpoint}

\end{document}